# On-Line Portfolio Selection with Moving Average Reversion

Bin Li                                                                                  S080061@NTU.EDU.SG
Steven C. H. Hoi                                                                        CHHOI@NTU.EDU.SG
School of Computer Engineering, Nanyang Technological University, Singapore 639798

## Abstract

On-line portfolio selection has attracted increasing interests in machine learning and AI communities recently. Empirical evidence show that stock's high and low prices are temporary and stock price relatives are likely to follow the mean reversion phenomenon. While the existing mean reversion strategies are shown to achieve good empirical performance on many real datasets, they often make the *single-period mean reversion* assumption, which is not always satisfied, leading to poor performance in some real datasets. To overcome the limitation, this article proposes a *multiple-period mean reversion*, or so-called "Moving Average Reversion" (MAR), and a new on-line portfolio selection strategy named "On-Line Moving Average Reversion" (OLMAR), which exploits MAR by applying powerful online learning techniques. From our empirical results, we found that OLMAR can overcome the drawbacks of existing mean reversion algorithms and achieve significantly better results, especially on the datasets where existing mean reversion algorithms failed. In addition to superior performance, OLMAR also runs extremely fast, further supporting its practical applicability to a wide range of applications.

## 1. Introduction

Portfolio selection, which has been explored in both finance and quantitative fields, aims to obtain certain targets in the long run by sequentially allocating the wealth among a set of assets. Mean-variance theory (Markowitz, 1952), which trades off between the expected return (mean) and risk (variance) of a portfolio, is suitable for single period portfolio selection. Contrarily, Kelly investment (Kelly, 1956; Breiman, 1961; Finkelstein & Whitley, 1981), which maximizes the expected log return of a portfolio, aims for multiple periods portfolio selection. Due to the sequential nature, recent on-line portfolio selection techniques often design algorithms following the second approach.

One important property exploited by many existing studies (Borodin et al., 2004; Li et al., 2011b; 2012) is the mean reversion property, which assumes poor performing stocks will perform well in the subsequent periods and vice versa, may better fit the financial markets. Though some recent mean reversion algorithms (Li et al., 2011b; 2012) achieve the best results on many datasets, they perform extremely poor on certain datasets, such as DJA dataset (Borodin et al., 2004). Comparing with Borodin et al. (2004), which exploits multi-period correlation, we found that the assumption of single-period prediction may attribute to the performance degradation. On the other hand, as illustrated in existing studies (Li et al., 2011b; 2012), Borodin et al. (2004), which heuristically exploits mean reversion via correlation, cannot fully exploit the potential of (multi-period) mean reversion.

To address the above drawbacks, we present a new approach for on-line portfolio selection, named "On-Line Moving Average Reversion" (OLMAR). The basic idea is to represent multi-period mean reversion as "Moving Average Reversion" (MAR), which explicitly predicts next price relatives using moving averages, and then learn portfolios by online learning techniques. To the best of our knowledge, OLMAR is the first algorithm that exploits moving average in the setting of on-line portfolio selection. Though simple in nature, OLMAR has a reasonable update and is empirically validated via extensive experiments on real markets. The experiments show that OLMAR can achieve better performance (in terms of *cumulative wealth*) than existing algorithms, and more importantly, can avoid the performance degradation in certain datasets, such as DJA dataset (Borodin et al., 2004; Li et al., 2012). Finally,

---

OLMAR runs much faster than the state of the art, and thus is suitable for large-scale applications.

The rest of the paper is organized as follows. Section 2 formulates the on-line portfolio selection problem, and Section 3 reviews and analyzes related work. Section 4 presents the proposed OLMAR approach, and its effectiveness is validated by extensive empirical studies on real stock markets in Section 5. Section 6 summarizes the paper and provides directions for future work.

## 2. Problem Setting

Consider an investment task over a financial market with $m$ assets for $n$ periods. On the $t^{\text{th}}$ period, the assets' prices are represented by a *close price vector* $\mathbf{p}_t \in \mathbb{R}_+^m$, and each element $p_{t,i}$ represents the close price of asset $i$. Their price changes are represented by a *price relative vector* $\mathbf{x}_t \in \mathbb{R}_+^m$, and $x_{t,i} = \frac{p_{t,i}}{p_{t-1,i}}$. Thus, an investment in asset $i$ on the $t^{\text{th}}$ period increases by a factor of $x_{t,i}$. Let us denote $\mathbf{x}^n = \{\mathbf{x}_1, \ldots, \mathbf{x}_n\}$ as the sequence of price relative vectors for $n$ periods.

An investment on the $t^{\text{th}}$ period is specified by a *portfolio vector* $\mathbf{b}_t = (b_{t,1}, \ldots, b_{t,m})$, where $b_{t,i}$ represents the proportion of wealth invested in asset $i$. Typically, we assume the portfolio is self-financed and no margin/short is allowed, therefore each entry of a portfolio is non-negative and adds up to one, that is, $\mathbf{b}_t \in \Delta_m$, where $\Delta_m = \{\mathbf{b}_t : \mathbf{b}_t \in \mathbb{R}_+^m, \sum_{i=1}^m b_{t,i} = 1\}$. The investment procedure is represented by a *portfolio strategy*, that is, $\mathbf{b}_1 = \frac{1}{m}\mathbf{1}$ and following sequence of mappings $\mathbf{b}_t : \mathbb{R}_+^{m(t-1)} \to \Delta_m, t = 2, 3, \ldots$, where $\mathbf{b}_t = \mathbf{b}_t(\mathbf{x}^{t-1})$ is the $t^{\text{th}}$ portfolio given past market sequence $\mathbf{x}^{t-1} = \{\mathbf{x}_1, \ldots, \mathbf{x}_{t-1}\}$. We denote by $\mathbf{b}^n = \{\mathbf{b}_1, \ldots, \mathbf{b}_n\}$ the strategy for $n$ periods.

On the $t^{\text{th}}$ period, a portfolio $\mathbf{b}_t$ produces a *portfolio period return* $s_t$, that is, the wealth increases by a factor of $s_t = \mathbf{b}_t^\top \mathbf{x}_t = \sum_{i=1}^m b_{t,i} x_{t,i}$. Since we reinvest and adopt price relative, the portfolio wealth would multiplicatively grow. Thus, after $n$ periods, a portfolio strategy $\mathbf{b}^n$ produces a *portfolio cumulative wealth* of $S_n$, which increases the initial wealth by a factor of $\prod_{t=1}^n \mathbf{b}_t^\top \mathbf{x}_t$, that is, $S_n(\mathbf{b}^n, \mathbf{x}^n) = S_0 \prod_{t=1}^n \mathbf{b}_t^\top \mathbf{x}_t$, where $S_0$ is set to \$1 for convenience.

Finally, let us formulate the on-line portfolio selection problem. In this task, a portfolio manager is a decision maker, whose goal is to produce a portfolio strategy $\mathbf{b}^n$, aiming to maximize the cumulative wealth $S_n$. He/she computes the portfolios sequentially. On each period $t$, the manager has access to the sequence of previous price relative vectors $\mathbf{x}^{t-1}$. Then, he/she computes a new portfolio $\mathbf{b}_t$ for next price relative vector $\mathbf{x}_t$, where the decision criterion varies among different managers. The portfolio $\mathbf{b}_t$ is scored based on portfolio period return $s_t$. This procedure is repeated until the end, and the portfolio strategy is finally scored according to portfolio cumulative wealth $S_n$.

Note that the above model in general assumes zero transaction cost/tax, perfect market liquidity, and zero impact cost. These assumptions are not trivial, and their effects and implications will be further analyzed and discussed in Section 5.3 and Section 5.5.

## 3. Related Work

The research of on-line portfolio selection grounds on the principle of Kelly investment (Kelly, 1956; Breiman, 1961; Thorp, 1971; Finkelstein & Whitley, 1981), that is, to maximize the expected *log* return of a portfolio. One classical strategy is *Constant Rebalanced Portfolios* (CRP), which follows Kelly's idea of keeping a fixed fraction for each asset on all periods. The best possible CRP strategy in hindsight is often known as *Best Constant Rebalanced Portfolios* (BCRP), which is the optimal strategy if the market is i.i.d. (Cover & Thomas, 1991, Theorem 15.3.1). Cover (1991) initialized the research of on-line portfolio selection (Ordentlich & Cover, 1996) and proposed *Universal Portfolios* (UP) strategy, whose portfolio is historical performance weighted average of all CRPs.

Helmbold et al. (1998) proposed *Exponential Gradient* (EG) strategy, which maximizes the expected log portfolio return estimated by last price relatives, and minimizes the deviation from last portfolio. Gaivoronski & Stella (2000) proposed *Successive Constant Rebalanced Portfolios* (SCRP), which maximizes the expected log cumulative wealth estimated using all historical price relative relatives[1]. Agarwal et al. (2006) proposed *Online Newton Step* (ONS), which extends the idea of SCRP by appending a regularization term to minimize the variation of next portfolio.

Borodin et al. (2004) proposed *Anti-correlation* (Anticor), which bets on the consistency of positive lagged cross-correlation and negative auto-correlation. Li et al. (2012) proposed *Passive Aggressive Mean Reversion* (PAMR), which iteratively chooses portfolio minimizing the expected return based on last price relatives. Li et al. (2011b) proposed *Confidence Weighted Mean Reversion* (CWMR) strategy, which exploits the mean reversion property and the variance information of portfolio.

Györfi et al. (2006) introduced the framework of non-parametric investment strategies, which searches over historical market sequence and identify a sample set of vectors, whose previous market windows are similar to recent window, and obtains a BCRP portfolio based on the set. With this framework, *Nonparametric Kernel-based moving window* (B$^\text{K}$) (Györfi et al., 2006) measures the similarity using Euclidean distance. Further, *Nonparametric Nearest Neighbor* (B$^\text{NN}$) (Györfi et al., 2008) searches for $\ell$ nearest neighbors to the recent market window. Recently, Li et al. (2011a) proposed *Correlation-driven Nonparametric learning* (CORN) to search for similar vectors via correlation.

---

[1] This idea also appeared in Ordentlich (1996, Chap. 4).

*Table 1.* Summary of the existing optimization formulations and their underlying predictions. $R(\cdot)$ denotes the regularization term, such as $L_2$ norm. $\text{Prob}(\cdot)$ denotes a probability function. PAMR/CWMR's prediction is not an strict equivalence, which we do not proof.

| Categories | Methods | Formulations | Prediction ($\tilde{\mathbf{x}}_{t+1}^i$) | Prob. ($p_i$) |
|---|---|---|---|---|
| In hindsight | BCRP | $\mathbf{b}_{t+1} = \arg\max_{\mathbf{b} \in \Delta_m} \sum_{i=1}^{n} \frac{1}{n} \log \mathbf{b} \cdot \mathbf{x}_i$ | $\mathbf{x}_i, i=1,\ldots,n$ | $1/n$ |
| 1 | EG | $\mathbf{b}_{t+1} = \arg\max_{\mathbf{b} \in \Delta_m} \log \mathbf{b} \cdot \mathbf{x}_t - \lambda R(\mathbf{b}, \mathbf{b}_t)$ | $\mathbf{x}_t$ | 1.00 |
| 1 | PAMR | $\mathbf{b}_{t+1} = \arg\min_{\mathbf{b} \in \Delta_m} \mathbf{b} \cdot \mathbf{x}_t + \lambda R(\mathbf{b}, \mathbf{b}_t)$ | $1/\mathbf{x}_t$ | 1.00 |
| 1 | CWMR | $\mathbf{b}_{t+1} = \arg\min_{\mathbf{b} \in \Delta_m} \text{Prob}(\mathbf{b} \cdot \mathbf{x}_t) + \lambda R(\mathbf{b}, \mathbf{b}_t)$ | $1/\mathbf{x}_t$ | 1.00 |
| 2 | SCRP | $\mathbf{b}_{t+1} = \arg\max_{\mathbf{b} \in \Delta_m} \sum_{i=1}^{t} \frac{1}{t} \log \mathbf{b} \cdot \mathbf{x}_i$ | $\mathbf{x}_i, i=1,\ldots,t$ | $1/t$ |
| 2 | ONS | $\mathbf{b}_{t+1} = \arg\max_{\mathbf{b} \in \Delta_m} \sum_{i=1}^{t} \frac{1}{t} \log \mathbf{b} \cdot \mathbf{x}_i - \lambda R(\mathbf{b})$ | $\mathbf{x}_i, i=1,\ldots,t$ | $1/t$ |
| 3 | B$^\text{K}$/B$^\text{NN}$/CORN | $\mathbf{b}_{t+1} = \arg\max_{\mathbf{b} \in \Delta_m} \sum_{i \in C_t} \frac{1}{|C_t|} \log \mathbf{b} \cdot \mathbf{x}_i$ | $\mathbf{x}_i, i \in C_t$ | $1/|C_t|$ |

### 3.1. Analysis of Existing Work

Most existing formulations follow the basic routine of Kelly-based portfolio selection (Thorp, 1971). In particular, a portfolio manager first predicts $\tilde{\mathbf{x}}_{t+1}$ in terms of $k$ possible values $\tilde{\mathbf{x}}_{t+1}^1, \ldots, \tilde{\mathbf{x}}_{t+1}^k$ and their corresponding probabilities $p_1, \ldots, p_k$. Note that each $\tilde{\mathbf{x}}_{t+1}^i$ denotes one possible combination vector of individual price relative prediction. Then he/she can figure out portfolio by maximizing the expected log return,

$$\mathbf{b}_{t+1} = \arg\max_{\mathbf{b} \in \Delta_m} \sum_{i=1}^{k} p_i \log\left(\mathbf{b} \cdot \tilde{\mathbf{x}}_{t+1}^i\right).$$

Based on methods of predicting $\tilde{\mathbf{x}}_{t+1}^i$ and $p_i$, most existing algorithms can be classified into three categories. Their optimization formulations and underlying predictions are summarized in Table 1, and their details can be found on their respective studies. Note that we have transformed certain formulations, however, maintained their key ideas.

Now let us focus on the first category, which consists of EG, PAMR and CWMR. Algorithms in this category assume a single prediction value with a probability of 100%, and maintains previous portfolio information via a regularization techniques. In particular, EG assumes $\tilde{\mathbf{x}}_{t+1}^1 = \mathbf{x}_t$ with $p_1 = 100\%$, while PAMR and CWMR assume $\tilde{\mathbf{x}}_{t+1}^1 = \frac{1}{\mathbf{x}_t}$ with $p_1 = 100\%$, which is in essence mean reversion idea. Note that the formulations of PAMR and CWMR ignore the log utility due to the single-value prediction and the consideration of convexity and computation issue. Though all three algorithms assume that all information is fully reflected by $\mathbf{x}_t$, their performance diverges and supports that mean reversion fits the markets. On the one hand, even with a decent theoretical result, EG always performs far behind. On the other hand, though without theoretical guarantees, PAMR and CWMR always produce the best results in various real markets. However, PAMR and CWMR suffer from dramatic failures when such single-period mean reversion is not satisfied (Li et al., 2012), which motivates our approach.

## 4. On-Line Moving Average Reversion

### 4.1. Motivation

Empirical results (Li et al., 2011b; 2012) show that mean reversion, which assumes the poor stock may perform good in the subsequent periods, may better fit the markets. PAMR and CWMR can exploit the mean reversion property well and achieve good results on most datasets at the time, especially the NYSE benchmark dataset (Cover, 1991). However, they rely on a simple assumption that the predicted next price relative $\tilde{\mathbf{x}}_{t+1}$ will be inverse proportion to last price relative $\mathbf{x}_t$. In particular, they implicitly assume that next price $\tilde{\mathbf{p}}_{t+1}$ will revert to last price $\mathbf{p}_{t-1}$, as follows,

$$\tilde{\mathbf{x}}_{t+1} = \frac{1}{\mathbf{x}_t} \Longrightarrow \frac{\tilde{\mathbf{p}}_{t+1}}{\mathbf{p}_t} = \frac{\mathbf{p}_{t-1}}{\mathbf{p}_t} \Longrightarrow \tilde{\mathbf{p}}_{t+1} = \mathbf{p}_{t-1}.$$

Note that $\mathbf{x}$ and $\mathbf{p}$ are all vectors and the above operations are element-wise.

Though empirically effective on most datasets, PAMR and CWMR's *single-period* assumption causes two potential problems. Firstly, both algorithms suffer from the frequently fluctuating raw prices, as they often contain a lot of noises. Second, their assumption of single-period mean reversion may not be satisfied in the real world. Even two consecutive declining price relatives, which are common, can fail both algorithms. One real example (Li et al., 2012) is DJA dataset (Borodin et al., 2004), on which PAMR performs the worst among the state of the art. Thus,

On-Line Portfolio Selection with Moving Average ReversionnoOn-Line Portfolio Selection with Moving Average Reversion*Table 2.* Illustration of growth of mean reversion strategies on toy markets. OLMAR is calculated with window size 2.

| Market Sequences | BCRP | PAMR/CWMR | OLMAR |
|---|---|---|---|
| $A: (1,2),\left(1,\frac{1}{2}\right),(1,2),\left(1,\frac{1}{2}\right),\ldots$ | $\left(\frac{9}{8}\right)^{n/2}$ | $\frac{3}{2}\times 2^{\frac{n-1}{2}}$ | $\frac{3}{2}\times \left(\frac{1}{2}\right)^{\frac{n-1}{2}}$ |
| $B: (1,2),(1,2),\left(1,\frac{1}{2}\right),\left(1,\frac{1}{2}\right),(1,2),\ldots$ | $\left(\frac{9}{8}\right)^{n/2}$ | $\frac{3}{2}$ | $\frac{3}{2}$ |
| $C: (1,2),(1,2),(1,2),\left(1,\frac{1}{2}\right),\left(1,\frac{1}{2}\right),\left(1,\frac{1}{2}\right),(1,2),\ldots$ | $\left(\frac{9}{8}\right)^{n/2}$ | $\frac{3}{2}\times \left(\frac{1}{2}\right)^{\frac{n-1}{6}}$ | $\frac{3}{2}\times 2^{\frac{n-1}{6}}$ |
| $D: \underbrace{(1,2),\ldots,(1,2)}_{k},\underbrace{(1,1/2),\ldots,(1,1/2)}_{k},(1,2),\ldots$ | $\left(\frac{9}{8}\right)^{n/2}$ | $\frac{3}{2}\times \left(\frac{1}{2}\right)^{(n-1)\times\left(\frac{1}{2}-\frac{1}{k}\right)}$ | $\frac{3}{2}\times \left(\frac{1}{2}\right)^{(n-1)\times\left(\frac{1}{k}-\frac{1}{2}\right)}$ |

traders are more likely to predict prices using the long-term mean. Also on the DJA dataset, Anticor, which exploits the *multi-period* statistical correlation, performs much better. However, as illustrated in Li et al. (2011b; 2012), due to its heuristic nature, Anticor can not fully exploit mean reversion. The two problems caused by single-period assumption and Anticor's inability to fully exploit mean reversion call for a more powerful approach to effectively exploit mean reversion, especially in terms of multi-period.

Now let us see the classic example (Cover & Gluss, 1986; Li et al., 2012) to illustrate the drawbacks of single-period mean reversion, as shown in Table 2. The toy market consists of cash and one volatile stock, whose market sequence follows A. It is easy to prove that BCRP ($\mathbf{b}=\left(\frac{1}{2},\frac{1}{2}\right)$) can grow by a factor of $\left(\frac{9}{8}\right)^{n/2}$, while PAMR/CWMR can grow by a better factor of $\frac{3}{2}\times 2^{(n-1)/2}$. Note that this virtual sequence is essentially singe-period mean reversion, which perfectly fits with PAMR and CWMR's assumption. However, if market sequence does not satisfy this assumption, both PAMR and CWMR would fail badly. Let us extend the market sequence to a two-period reversion, that is, market sequence B. In such a market, BCRP can achieve the same growth as before. Contrarily, PAMR/CWMR can achieve a constant wealth $\frac{3}{2}$, which has no growth! More generally, if we further extend to $k$-period mean reversion, BCRP can still achieve the same growth, while PAMR/CWMR will grow to $\frac{3}{2}\times \left(\frac{1}{2}\right)^{(n-1)\times\left(\frac{1}{2}-\frac{1}{k}\right)}$, which definitely approaches bankruptcy when $k\geq 3$.

To better exploit the (multi-period) mean reversion property, we proposed a new type of algorithms, named *On-Line Moving Average Reversion* (OLMAR), for on-line portfolio selection. The essential idea is to exploit multi-period moving average (mean) reversion via power online machine learning. Rather than $\tilde{\mathbf{p}}_{t+1}=\mathbf{p}_{t-1}$, OLMAR assumes that next price will revert to *Moving Average* (MA) at the end of $t^{th}$ period, that is, $\tilde{\mathbf{p}}_{t+1}=MA_t(w)=\frac{1}{w}\sum_{i=t-w+1}^{i=t}\mathbf{p}_i$, where $MA_t(w)$ denotes the *Moving Average* (MA) with a $w$-window. In time series analysis, MA is typically used to smooth short-term price fluctuations and focus on long-term trends, thus can solve the two drawbacks of existing mean reversion algorithms. To this end, we propose a price relative vector following the idea of "Moving Average Reversion" (MAR),

$$\tilde{\mathbf{x}}_{t+1}(w)=\frac{MA_t(w)}{\mathbf{p}_t}=\frac{1}{w}\left(\frac{\mathbf{p}_t}{\mathbf{p}_t}+\frac{\mathbf{p}_{t-1}}{\mathbf{p}_t}+\cdots+\frac{\mathbf{p}_{t-w+1}}{\mathbf{p}_t}\right)$$
$$=\frac{1}{w}\left(1+\frac{1}{\mathbf{x}_t}+\cdots+\frac{1}{\bigotimes_{i=0}^{w-2}\mathbf{x}_{t-i}}\right), \quad (1)$$

where $w$ is the window size and $\bigotimes$ denotes element-wise production. Without detailing the calculation, we list the growth of OLMAR in different toy markets in Table 2. Clearly, OLMAR performs much better in the scenarios of multi-period mean reversion, while performs poor in single-period reversion. Our further empirical analysis shows that the markets are more likely to follow multi-period reversion.

Based on the expected price relative vector in Eq. (1), OLMAR further adopts the idea of an effective online learning algorithm, that is, Passive Aggressive (PA) (Crammer et al., 2006) learning, to exploit moving average reversion. Generally proposed for classification, PA passively keeps the previous solution if the classification is correct, while aggressively approaches a new solution if the classification is incorrect. After formulating the proposed OLMAR optimization, we solve its closed form update and design the proposed algorithm.

### 4.2. Formulation

The proposed formulation, OLMAR, is to exploit moving average reversion via PA online learning. The basic idea is to maximize the expected return $\mathbf{b}\cdot\tilde{\mathbf{x}}_{t+1}$, and keep the last portfolio information via regularization term. Thus, we follow the similar idea of PAMR (Li et al., 2012) and formulate an optimization,

**Optimization Problem: OLMAR**

$$\mathbf{b}_{t+1}=\operatorname*{arg\,min}_{\mathbf{b}\in\Delta_m}\quad \frac{1}{2}\|\mathbf{b}-\mathbf{b}_t\|^2 \quad \text{s.t.}\quad \mathbf{b}\cdot\tilde{\mathbf{x}}_{t+1}\geq\epsilon.$$

Note that we adopt expected return rather than expected log return. According to Helmbold et al.

**Algorithm 1** Portfolio Selection with OLMAR.
1: **Input:** $\epsilon > 1$: Reversion threshold; $w \geq 3$: Window size; $\mathbf{x}_1^n$: Market sequence;
2: **Output:** $S_n$: Cumulative wealth after $n^{th}$ periods
3: **Procedure:**
4: Initialization: $\mathbf{b}_1 = \frac{1}{m}\mathbf{1}$, $S_0 = 1$;
5: **for** $t = 1, 2, \ldots, n$ **do**
6:    Receive stock price relatives: $\mathbf{x}_t$
7:    Calculate daily return and cumulative return:
$S_t = S_{t-1} \times (\mathbf{b}_t \cdot \mathbf{x}_t)$
8:    Predict next price relative vector:
$$\tilde{\mathbf{x}}_{t+1}(w) = \frac{1}{w}\left(1 + \frac{1}{\mathbf{x}_t} + \cdots + \frac{1}{\bigotimes_{i=0}^{w-2} \mathbf{x}_{t-i}}\right)$$
9:    Update the portfolio:
$\mathbf{b}_{t+1} = \text{OLMAR}(\epsilon, w, \tilde{\mathbf{x}}_{t+1}, \mathbf{b}_t)$
10: **end for**

**Algorithm 2** OLMAR$(\epsilon, w, \tilde{\mathbf{x}}_{t+1}, \mathbf{b}_t)$.
1: **Input:** $\epsilon > 1$: Reversion threshold; $w \geq 3$: Window size; $\tilde{\mathbf{x}}_{t+1}$: Predicted price relatives; $\mathbf{b}_t$: Current portfolio;
2: **Output:** $\mathbf{b}_{t+1}$: Next portfolio;
3: **Procedure:**
4: Calculate the following variables:
$$\bar{x}_{t+1} = \frac{\mathbf{1}^\top \tilde{\mathbf{x}}_{t+1}}{m}, \lambda_{t+1} = \max\left\{0, \frac{\epsilon - \mathbf{b}_t \cdot \tilde{\mathbf{x}}_{t+1}}{\|\tilde{\mathbf{x}}_{t+1} - \bar{x}_{t+1}\mathbf{1}\|^2}\right\}$$
5: Update the portfolio:
$\mathbf{b}_{t+1} = \mathbf{b}_t + \lambda_{t+1}(\tilde{\mathbf{x}}_{t+1} - \bar{x}_{t+1}\mathbf{1})$
6: Normalize $\mathbf{b}_{t+1}$:
$$\mathbf{b}_{t+1} = \arg\min_{\mathbf{b} \in \Delta_m} \|\mathbf{b} - \mathbf{b}_{t+1}\|^2$$

(1998), to solve the optimization with expected log return, one can adopt the first-order Taylor expansion, which is essentially linear.

The above formulation explicitly reflects the basic idea of the proposed OLMAR. On the one hand, if the constraint is satisfied, that is, the expected return is higher than a threshold, then the resulting portfolio equals to the previous portfolio. On the other hand, if the constraint is not satisfied, then the formulation will figure out a new portfolio such that the expected return is higher than the threshold, while the new portfolio is not far from the last one.

Since OLMAR follows the same learning principle as PAMR, their formulations are similar. However, the two solutions are essentially different. In particular, PAMR's core constraint, i.e., $\mathbf{b} \cdot \mathbf{x}_t \leq \epsilon$, adopts the raw price relative and has a different inequality sign.

### 4.3. Algorithm

The above formulation is thus convex and straightforward to solve via convex optimization (Boyd & Vandenberghe, 2004). We now derive the OLMAR solution as illustrated in Proposition 1.

**Proposition 1.** *The solution of OLMAR without considering the non-negativity constraint is*
$$\mathbf{b}_{t+1} = \mathbf{b}_t + \lambda_{t+1}(\tilde{\mathbf{x}}_{t+1} - \bar{x}_{t+1}\mathbf{1}),$$
*where $\bar{x}_{t+1} = \frac{1}{m}(\mathbf{1} \cdot \tilde{\mathbf{x}}_{t+1})$ denotes the average predicted price relative and $\lambda_{t+1}$ is the Lagrangian multiplier calculated as,*
$$\lambda_{t+1} = \max\left\{0, \frac{\epsilon - \mathbf{b}_t \cdot \tilde{\mathbf{x}}_{t+1}}{\|\tilde{\mathbf{x}}_{t+1} - \bar{x}_{t+1}\mathbf{1}\|^2}\right\}.$$

It is worth noting that in the derivation we do not consider the non-negativity constraint following Helmbold et al. (1998). Thus, it is possible that the resulting portfolio goes out the portfolio simplex domain. To maintain a proper portfolio, we finally project the portfolio calculated according to Proposition 1 to the simplex domain.

To this end, we can design the proposed algorithm based on the proposition. The on-line portfolio selection following the problem setting in Section 2 is illustrated in Algorithm 1, and the proposed OLMAR procedure is illustrated in Algorithm 2.

Empirical observations of the parameter sensitivity in Section 5.2 show that the final performance is sensitive to the parameter $w$. To smooth the final performance of the system, we propose to adopt the Buy and Hold (BAH) version (Borodin et al., 2004; Györfi et al., 2006; Li et al., 2012), that is, for each period, we treat the individual OLMAR with a specified $w \geq 3$ as an expert and combine multiple experts' portfolios weighted by their historical performance. We denote the the algorithm as BAH$_W$(OLMAR) with a parameter $W$ denoting the maximum window size, that is, BAH$_W$(OLMAR) combines $W-2$ individual OLMAR experts with $w = 3, \ldots, W$.

### 4.4. Analysis

The update of OLMAR is straightforward, that is, $\mathbf{b}_{t+1} = \mathbf{b}_t + \lambda_{t+1}(\tilde{\mathbf{x}}_{t+1} - \bar{x}_{t+1}\mathbf{1})$. Basically, the update divides the assets into two groups by prediction average. For the assets in the group with higher predictions than average, OLMAR increases their proportions; for the others, OLMAR decreases their proportions. The transferred proportions are related to the surprise of the predictions over their average and the positive Lagrangian multiplier, $\lambda$. This is consistent with normal portfolio selection procedure, that is, to transfer the wealth to the assets with better prospect to grow.

Clearly, the OLMAR update costs linear time per period w.r.t. $m$, and the normalization step can also be implemented in linear time (Duchi et al., 2008). To the best of our knowledge, OLMAR's linear time is no worse than any existing algorithm.

## 5. Experiments

We now evaluate the effectiveness of the proposed OLMAR algorithm by performing an extensive set of experiments on publicly available and diverse datasets from real stock markets.

Table 3 details the four experimental datasets[2]. Among these, NYSE (O) is one benchmarks dataset evaluated by all existing methods, while NYSE (N) is a dataset collected by Li et al. (2011b) as a continuation of NYSE (O). The remaining two datasets (DJA and TSE) are collected by Borodin et al. (2004).

Table 3. Summary of 4 real datasets from various markets.

| Dataset | Region | Time frame | # days | # assets |
| --- | --- | --- | --- | --- |
| NYSE (O) | US | 3/7/1962 - 31/12/1984 | 5651 | 36 |
| NYSE (N) | US | 1/1/1985 - 30/6/2010 | 6431 | 23 |
| DJA | US | 1/1/2001 - 14/1/2003 | 507 | 30 |
| TSE | CA | 4/1/1994 - 31/12/1998 | 1259 | 88 |

In this paper, we adopt the most common metric, *cumulative wealth*, to measure the investment performance. Results with other metrics, including *risk-adjusted return*, will be included in the long version of this paper. We compare the proposed approach with all existing methods in Section 3, whose parameters are set according to their respective studies.

In our experiments, we implement the proposed OLMAR, and its BAH version BAH(OLMAR). In all cases, we empirically set the parameters, that is, $\epsilon = 10$ and $w = 5$, which provide a consistent results for OLMAR in all cases. For BAH(OLMAR), we set their maximum window size $W = 30$, resulting in 28 OLMAR experts with $w = 3$ to 30. In addition to the empirical selection of parameters, we also evaluate the parameter scalability of the two possible parameters in Section 5.2. Finally, we also refer the best performance (in hindsight) among the underlying experts of the BAH version, named MAX(OLMAR).

### 5.1. Cumulative Wealth

Figure 1(a) illustrates the main results of this study, that is, the cumulative wealth achieved by various approaches. The results clearly show that OLMAR achieves the top performance among all competitors. On the well-known benchmark NYSE (O) dataset, OLMAR significantly outperforms the state of the art; the similar observation is also found on the successive NYSE (N) dataset. Besides, though most existing algorithms except Anticor perform bad on the DJA dataset, OLMAR achieves the best performance. Moreover, the maximum performance show that it is possible to achieve better performance via effective expert combination. Finally, the $t$-test statistics (Grinold & Kahn, 1999) shown in Table 1(b) validate that the results achieved are not due to luck.

### 5.2. Parameter Sensitivity

Now let us evaluate algorithm's sensitivity to its parameters, that is, $\epsilon$ and $w$. Figure 2 shows the sensitivity of $\epsilon$ with fixed $w = 5$ and Figure 3 show the sensitivity of $w$ with fixed $\epsilon = 10$. From the former, we can observe that in general the total wealth sharply increases when $\epsilon$ approaches 1 and flattens when $\epsilon$ crosses a threshold. From the latter, we can observe that as $w$ increases, the performance initially increases, spikes with a data-dependant value, and then decreases. Anyway, its performance with most choices of $\epsilon$ and $w$ have a much better performance than BCRP strategy and the market. Moreover, the latter figure also show that the Buy and Hold versions greatly smooth the performance with varying $w$ of the underlying experts. All above observations also show that it is robust to the choices of parameters and is convenient to choose satisfying parameters.

### 5.3. Transaction Cost Scalability

To evaluate the practical applicability, we evaluate the scalability of the proposed algorithms with respect to proportional transaction cost (Borodin et al., 2004). Figure 4 illustrates the cumulative wealth achieved by OLMAR with increasing transaction cost rate $\gamma$, and also the results obtained by four representative algorithms (two benchmarks and two state of the arts in different categories). On the one hand, the results clearly show that OLMAR can withstand reasonable transaction cost rates, as it often has high break-even rates with respect to the market. On the other hand, OLMAR can outperform the benchmarks and state of the arts, under various transaction cost rates. In a word, OLMAR performs excellent when trading is not frictionless, supporting its practical applicability.

### 5.4. Computational Time

Finally, we evaluate the computational time as shown in Table 1(c). Note that we only list the computational times of methods with comparable performance. Theoretically, as we analyzed in Section 4.4, OLMAR enjoys linear computational time complexity. Empiri-

---

[2] All datasets, including the composites, are available on http://www.cais.ntu.edu.sg/~libin/portfolios/.

On-Line Portfolio Selection with Moving Average Reversion

| Methods | NYSE (O) | NYSE (N) | DJA | TSE |
|---|---|---|---|---|
| Market | 14.50 | 18.06 | 0.76 | 1.61 |
| Best-stock | 54.14 | 83.51 | 1.19 | 6.28 |
| BCRP | 250.60 | 120.32 | 1.24 | 6.78 |
| UP | 26.68 | 31.49 | 0.81 | 1.60 |
| EG | 27.09 | 31.00 | 0.81 | 1.59 |
| ONS | 109.19 | 21.59 | 1.53 | 1.62 |
| $B^K$ | 1.08E+09 | 4.64E+03 | 0.68 | 1.62 |
| $B^{NN}$ | 3.35E+11 | 6.80E+04 | 0.88 | 2.27 |
| CORN | 1.48E+13 | 5.37E+05 | 0.84 | 3.56 |
| Anticor | 2.41E+08 | 6.21E+06 | 2.29 | 39.36 |
| PAMR | 5.14E+15 | 1.25E+06 | 0.68 | 264.86 |
| CWMR | 6.49E+15 | 1.41E+06 | 0.68 | 332.62 |
| OLMAR | **3.68E+16** | **2.54E+08** | 2.06 | **424.80** |
| BAH(OLMAR) | 2.27E+16 | 1.41E+08 | **2.38** | 172.11 |
| MAX(OLMAR) | 1.62E+17 | 3.95E+08 | 3.30 | 1.18E+03 |

(a) Cumulative Wealth

| Stat. Attr. | NYSE (O) | NYSE (N) | DJA | TSE |
|---|---|---|---|---|
| Size | 5651 | 6431 | 507 | 1259 |
| MER (OLMAR) | 0.0074 | 0.0036 | 0.0020 | 0.0061 |
| MER (Market) | 0.0005 | 0.0005 | -0.0004 | 0.0004 |
| $\alpha$ | 0.0068 | 0.0030 | 0.0025 | 0.0056 |
| $\beta$ | 1.2965 | 1.1768 | 1.2627 | 1.5320 |
| $t$-statistics | 15.2405 | 7.3704 | 2.1271 | 3.4583 |
| $p$-value | 0.0000 | 0.0000 | 0.0169 | 0.0003 |

(b) Statistical Test of OLMAR

| Methods | NYSE (O) | NYSE (N) | DJA | TSE |
|---|---|---|---|---|
| $B^{NN}$ | 4.93E+04 | 3.39E+04 | 1.28E+03 | 1.32E+03 |
| CORN | 8.78E+03 | 1.03E+04 | 172 | 1.59E+03 |
| Anticor | 2.57E+03 | 1.93E+03 | 175 | 2.15E+03 |
| PAMR | 8 | 7 | 0.5 | 2 |
| CWMR | 123 | 68 | 9 | 162 |
| OLMAR | 4.0 | 3.3 | 0.3 | 0.7 |

(c) Computational Time (seconds)

*Figure 1.* Performance evaluation: (a). Cumulative wealth achieved by various trading strategies on the four datasets. The best results (excluding the best experts at the bottom, which is in hindsight) in each dataset are highlighted in *bold*. (b). Statistical $t$-test of the performance achieved by OLMAR on the stock datasets. MER denotes "Mean Excess Return". (c). Computational times (seconds) on the four datasets achieved by the state of the art.

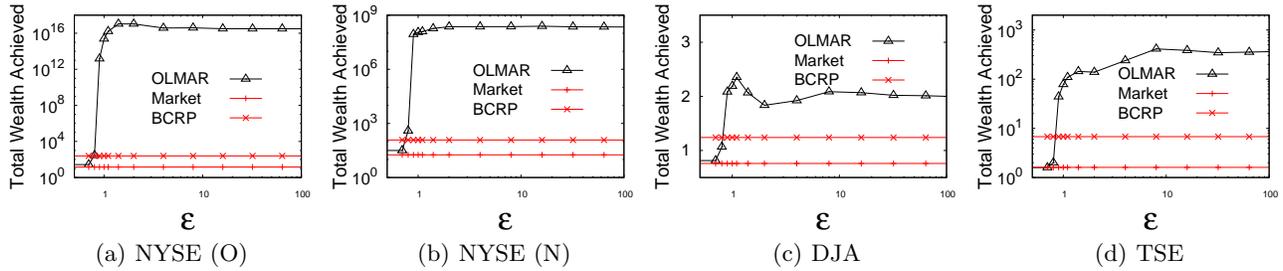

*Figure 2.* Parameter sensitivity of OLMAR w.r.t. $\epsilon$ with fixed $w$ ($w = 5$).

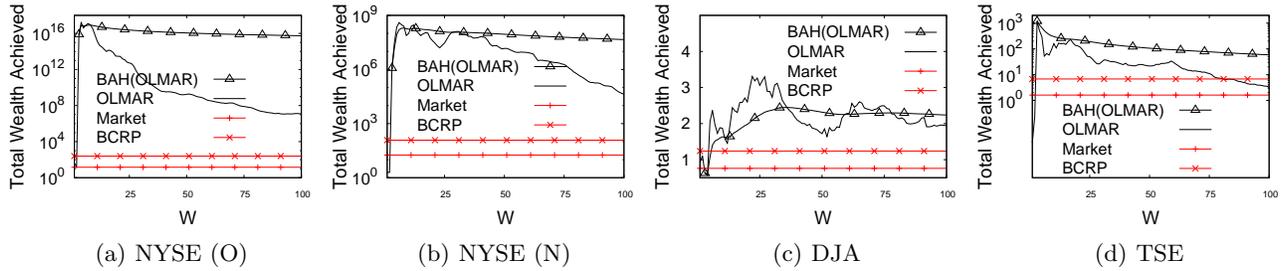

*Figure 3.* Parameter sensitivity of OLMAR w.r.t. $w$ with fixed $\epsilon$ ($\epsilon = 10$).

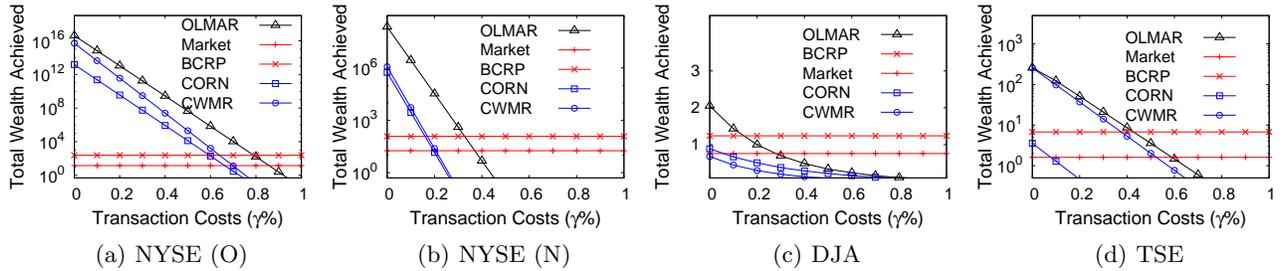

*Figure 4.* Scalability of the total wealth achieved by OLMAR with respect to transaction cost rate $\gamma\%$.

cally, as shown in the table, OLMAR algorithm takes the least times on all datasets. Note that with daily frequency, competitors' average times are acceptable, however, their times are not acceptable in the scenario of high frequency trading. Such time efficiency supports OLMAR's large-scale real applications.

### 5.5. Discussion and Thread of Validity

Without theoretical guarantee, the empirical assumptions in Section 2 are worth inspecting, as done by all existing heuristic algorithms (Borodin et al., 2004; Li et al., 2011a;b; 2012). As evaluated in Section 5.3, OLMAR can withstand reasonable rates of transaction cost. All the datasets are composed of the largest index composite stocks, which have the best market liquidity. To limit market impact when portfolio is too big, one solution is to scale-down the portfolio, as done by several quantitative funds. Here, we emphasize again that even in such a "perfect market", no algorithm has ever claimed such good performance.

There may also exist certain issues in the back tests. One possible issue is "survivorship bias". Since following existing works, the composition stocks never delisted from markets and survived for a long time, especially the two NYSE datasets. Another possible issue is "dataset selection". In fact, OLMAR also performs quite well on the other two public datasets (Li et al., 2012), that is, SP500 and MSCI, whose results will be included in the long version of this paper.

## 6. Conclusion

This paper proposes a novel online portfolio selection strategy named "On-Line Moving Average Reversion" (OLMAR), which exploits "Moving Average Reversion" via on-line learning algorithms. The approach can solve the problems of the state of the art caused by the single-period mean reversion and achieve satisfying results in real markets. It also runs extremely fast and is suitable for large-scale real applications. In future, we will further explore the theoretical aspect of the mean reversion property.

## Acknowledgments

This work was fully supported by Singapore MOE tier 1 project (RG33/11).